\documentclass[aps]{revtex4}
\usepackage{amssymb}
\usepackage{amsmath}
\usepackage{graphicx}
\usepackage{epsfig,amsmath}
\usepackage[bookmarksnumbered,linktocpage,pdfstartview=FitH]{hyperref}

\begin{document}

\title{Coarse-grained Hydrodynamics of turbulent superfluids: HVBK approach and the bundle structure of the  vortex tangle.}
\author{Sergey K. Nemirovskii\thanks{%
email address: nemir@itp.nsc.ru}}
\affiliation{Institute of Thermophysics, Lavrentyev ave., 1, 630090, Novosibirsk, Russia}

\begin{abstract}
In the paper I develop a critical analysis of the use of the HVBK method for
the study of three-dimensional turbulent flows of superfluids. The
conception of the vortex bundles forming the structure of quantum turbulence
is controversial and does not justify the use of the HVBK method. In
addition, this conception is counterproductive, because it gives incorrect
ideas about the structure of the vortex tangle as a set of bundles
containing parallel lines. The only type of dynamics of vortex filaments
inside these bundles is possible, namely, Kelvin waves running along the
filaments. At the same time, as shown in numerous numerical simulations, a
vortex tangle consists of a set of entangled vortex loops of different sizes
and having a random walk structure. These loops are subject to large
deformations (due to highly nonlinear dynamics), they reconnect with each
other and with the wall, split and merge, creating a lot of daughter loops.
They also bear Kelvin waves on them, but the latter have little impact.

I also propose and discuss an alternative variant of study of
three-dimensional turbulent flows, in which the vortex line density $%
\mathcal{L}(r,t)$ is not associated with $\nabla \times \mathbf{v}_{s}$, but
it is an independent equipollent variable described by a separate equation.
\end{abstract}

\maketitle


\section{Introduction}

The concept of quantum turbulence began with the well-known experiment by
Gorter and Mellink (GM)on heat exchange in superfluid helium and Feynman's
theory, which explains the GM experiment on the basis of the dynamics of a
set of chaotic quantized vortices, or vortex tangle \cite{Gorter1949},\cite%
{Feynman1955}. Since then, the concepts of quantum turbulence as a chaotic
tangle of vortex filaments have expanded and appeared in many fields of
physics, ranging from superfluids and cold atoms to heavy ions, and neutron
stars (see e.g. INT Program INT-19-1a, URL
http://www.int.washington.edu/PROGRAMS/19-1a/).

The experimental studies of superfluid turbulence is mainly based on
hydrodynamic methods. These include, for example, the study of the vortex
tangle using the first and second sounds, the measurement of the temperature
drop or the pressure field, the investigation of fluctuating velocity fields
(via studying pressure fluctuation), etc. On the other hand, most of the
known methods of creating superfluid turbulence also have the hydrodynamic
origin. These are the generation of a vortex tangle by a flow (counterflow),
by powerful sound fields, using oscillating devices etc. Therefore,
knowledge how to describe hydrodynamic processes in superfluids in the
presence of the vortex tangle is very actual and important problem.

The article\ discusses the problem of construction of the coarse-grained
hydrodynamics of turbulent flows of superfluids. In particular, I conduct a
critical analysis of the use of the HVBK (Hall-Vinen-Bekarevich-Khalatnikov)
method for the study of three-dimensional flows of superfluids. Indeed, this
approach, which relates the vortex , line density $\mathcal{L}(r,t)$ with
the coarse-garined vorticity $\nabla \times \mathbf{v}_{s}$ via the favor
Feynman rule, was initially elaborated and is only suitable for rotating
stationary cases. Meanwhile, at present, the use of this method for
three-dimensional unsteady flows is a widespread practice. Sometime this is
done without justification, sometimes they refer to the so-called vortex
bundle structure of quantum turbulence. The conception of the vortex bundles
forming the structure of quantum turbulence is also critically discussed in
the paper. I also propose an alternative variant in which the vortex line
density $\mathcal{L}(r,t)$ is not associated with $\nabla \times \mathbf{v}%
_{s}$, but it is an independent and equipollent variable described by a
separate equation. The structure of paper is following. The next, second
section is devoted to general formulation of coarse-grained hydrodynamics of
superfluids in presence of the vortex tangle. In the third Section the
problem of the use of the HVBK methods for rotating superfluids and in the
three -dimensional flows are discussed. An alternative variant of study of
three-dimensional turbulent flow is described in fourth Section.

\section{Coarse-grained Hydrodynamics of turbulent superfluids.}

In presence of the vortex filaments the two-fluid hydrodynamics of the
superfluid helium should be modernized and be represented as follows:%
\begin{equation}
\rho _{n}{\frac{\partial \mathbf{v}_{n}}{\partial t}}+\rho _{n}(\mathbf{v}%
_{n}\cdot \nabla )\mathbf{v}_{n}=-{\frac{\rho _{n}}{\rho }}\nabla p_{n}-\rho
_{s}S\nabla T+\mathbf{F}_{mf}+\eta \nabla ^{2}\mathbf{v}_{n},
\label{equa-Vn}
\end{equation}%
\begin{equation}
\rho _{s}{\frac{\partial \mathbf{v}_{s}}{\partial t}}+\rho _{s}(\mathbf{v}%
_{s}\cdot \nabla )\mathbf{v}_{s}=-{\frac{\rho _{s}}{\rho }}\nabla p_{s}+\rho
_{s}S\nabla T-\mathbf{F}_{mf}  \label{equa-Vs}
\end{equation}

We assume that the motion of both components is incompressible, $\nabla
\cdot \mathbf{v}_{n}=0$, $\nabla \cdot \mathbf{v}_{s}=0$. and where $\mathbf{%
v}_{n}$ and $\mathbf{v}_{s}$ are the coarse-grained velocity of the normal
and superfluid component (averaged over a small volume $\mathcal{V}$), $%
p_{n} $ and $p_{s}$ are the effective pressures acting on the normal and the
superfluid component ($\nabla p_{n}=\nabla p+(\rho _{s}/2)\nabla v_{ns}^{2}$
and $\nabla p_{s}=\nabla p-(\rho _{n}/2)\nabla v_{ns}^{2}$), $p$ is the
total pressure, $S$ is the entropy, $T$ is the absolute temperature, $\eta $
is the dynamic viscosity of the normal component, and $\mathbf{v}_{ns}=%
\mathbf{v}_{n}-\mathbf{v}_{s}$.The effects of the vortices on the two
components (normal and superfluid) are described by the friction force
exerted by the superfluid component on the normal component $\mathbf{F}_{mf}$
.When these forces are averaged over all vortices inside the small volume $%
\mathcal{V}$, then the following expression of $\mathbf{F}_{mf}$\ is
obtained (see e.g. \cite{Schwarz1988})

\begin{equation}
\mathbf{F}_{mf}=\mathcal{L}<\mathbf{f}_{MF}>=\alpha \rho _{s}\kappa \mathcal{%
L}\left\langle \mathbf{s^{\prime }}\times \lbrack \mathbf{s^{\prime }}\times
(\mathbf{v}_{ns}-\mathbf{v_{i}})]\right\rangle +\alpha ^{\prime }\rho
_{s}\kappa \mathcal{L}\left\langle \mathbf{s^{\prime }}\times (\mathbf{v}%
_{ns}-\mathbf{v_{i}})\right\rangle .  \label{Fns-media}
\end{equation}%
In this equation $\mathbf{s^{\prime }(}\xi \mathbf{)}$ is the tangent vector
along the vortex filaments $\mathbf{s(}\xi \mathbf{)}$, composing the vortex
tangle, $\alpha ,\alpha ^{\prime }$ are temperature-dependent dimensionless
mutual friction parameters, $\mathbf{v_{i}}$.is the self-induced velocity of
the vortex filament. Quantity $\mathcal{L}$ is the vortex line density,
averaging $\left\langle \cdot \right\rangle $ is performed over various
configurations of vortex filaments. Equations (\ref{equa-Vn}) - (\ref%
{Fns-media}) are coarse-grained equations, hence the inclusion of the
effects of the vortex lines requires a high vortex line density per unit
volume. These equations had been written and discussed for a long time, a
classic variant, close to the stated above can be found in the book \cite%
{Donnelly1991}.

In the written form these equations are common to any type of flow, such as
counterflow, flow past obstacles, acoustic waves, etc. The main difference
for different types of flows is the determination of the vortex line density
$\mathcal{L}$ entering into the expression for the macroscopic mutual
friction force (\ref{Fns-media}), and the choice of averaging method $%
\left\langle \cdot \right\rangle $.

Another question concerns the temperature range for applying equations (\ref%
{equa-Vn}) - (\ref{Fns-media}), In general, the quantum turbulence is the
chaotic dynamics of three strongly interacting nonlinear fields. These are
the motion of the normal and superfluid component and stochastic evolution
of a set of vortex filaments (vortex tangle).\textsl{\ }The type of
interaction depends of the temperature $T$\ (via mutual friction). For large
$T$\ the coupling is strong, both components are clumped and move together.
As a result, we got almost the one-fluid hydrodynamics. The case of very low
temperature is very interesting because it is ideal for testing the idea of
whether the dynamics of discrete vortices (quantized vortex lines) can
imitate classical turbulence. The study of superfluid turbulence for
intermediate temperatures is not suitable for this purpose (due to the
presence of a normal component). Quantum turbulence in superfluid helium
(for intermediate temperatures) is rather the separate problem, not
identical to classical turbulence. And it is precisely this case that
requires the study of two coupled Navier-Stokes Equations. Thus, we can say
that we study the intermediate temperature range.

The above equations (\ref{equa-Vn}) - (\ref{Fns-media}) are a point of
consensus among physicists. Disagreements begin with the question how to
fulfil averaging and to treat variable $\mathcal{L}$. Here we discuss two
main ways to perform these procedures. They are the
Hall-Vinen-Bekarevich-Khalatnikov (HVBK) model (see e.g., book by
Khalatnikov, \cite{Khalatnikov1965} 1965) and the Hydrodynamics of
superfluid turbulence model (Nemirovskii \& Lebedev, 1983 \cite%
{Nemirovskii1983},\cite{Nemirovskii1995})

\section{HVBK approach}

\subsection{HVBK approach for rotating superfluids}

The Hall-Weinen-Bekarevich-Khalatnikov (HVBK) model (see, for example, the
book Khalatnikov \cite{Khalatnikov1965} became the basis for the
mathematical formalism of the hydrodynamics of rotating superfluids. As is
well known (see \cite{Feynman1955} , in a vessel rotating with an angular
velocity $\Omega $, appears a regular array of vortex filaments with a
density $n=2\Omega /\kappa $. Such a distribution of vortices creates an
average coarse-grained superfluid velocity $\left\langle \mathbf{v}%
_{s}\right\rangle $, which satisfies the condition of solid body rotation $%
\left\langle \mathbf{v}_{s}\right\rangle =\mathbf{\Omega \times r}$. The
vorticity filed $\mathbf{\omega }$ is $\mathbf{\omega }=2\mathbf{\Omega }$.
Therefore, the density of the vortex filaments in this case can be related
to the vorticity field by the following relation:%
\begin{equation}
\nabla \times \mathbf{v}_{s}=\kappa \mathcal{L}  \label{rot via L}
\end{equation}

~~~Due to the smallness of the quantum of circulation $\kappa $, even a
relatively weak rotational speed produces a high density of vortex lines.
Thus, it is possible to construct \ a coarse-grained hydrodynamics of
hydrodynamic equations, which the average contribution of many individual
vortex lines and incorporate their contribution to the macroscopic
evolutionary equations for superfluid and normal He II velocities.

Combining (\ref{Fns-media}) with (\ref{rot via L}) one get expression for
average coarse-garined mutual friction $\mathbf{F}_{mf}^{(HVBK)}$

\begin{equation}
\mathbf{F}_{mf}^{(HVBK)}=\rho _{s}\alpha {\hat{\omega}}\times \lbrack
\mathbf{\omega }\times (\mathbf{v}-\tilde{\beta}\nabla \times \mathbf{\hat{%
\omega}})]+\rho _{s}\alpha ^{\prime }\mathbf{\omega }\times (\mathbf{v}-%
\tilde{\beta}\nabla \times \mathbf{\hat{\omega}}),  \label{Fns-HVBK}
\end{equation}%
where $\mathbf{\omega }=\nabla \times \mathbf{v}_{s}$ is the averaged
superfluid vorticity, $\hat{\omega}=\mathbf{\omega }/|\omega |$. \ Thus, the
question of elimination of the vortex line density is resolved with the use
of the Feynman rule, it allows to study various problems of coarse-grained
dynamics of rotating supefluids (see, e.g. book be Sonin \cite{Sonin2016}).

\subsection{HVBK approach for three -dimensional flows}

The HVBK model is a fruitfull and elegant approach, however \emph{it is
principally assigned for rotating superfluids}. Nevertheless this approach,
which uses ansatz $\nabla \times \mathbf{v}_{s}=\kappa \mathcal{L}$ (\ref%
{rot via L}),$\mathcal{\ }$and the force $\mathbf{F}_{mf}^{(HVBK)}$ (\ref%
{Fns-HVBK}) is widely used for numerical and analytical studies of
coarse-grained hydrodynamic problems of turbulent superfluid in three
-dimensional situations.

This approach seems to be unfounded. In my opinion, there is no way to apply
it to three-dimensional hydrodynamics. Anticipating the objections of
readers, I would like to discuss the usual arguments for using the HVBK
method. Probably the only argument is the conviction that the vortex tangle
consists of so-called vortex bundles, unifying many near parallel vortex
filaments.

There are few papers (see, e.g. \cite{Baggaley2012a}, \cite{Sasa2011}),
where the authors claim of the existence of the bundles. In fact, they only
demonstrated how, with the help of statistical analysis, one can get a small
polarization (prevailing of one direction over the other) in the vortex
tangle. But, firstly, it is just a statistical effect and, secondly, under
no circumstances this small polarization allows to use the Feynman rule (\ref%
{rot via L}), which is crucially needed in the HVBK equations. To use this
ansatz, it is necessary that all filaments in the vortex tangle are involved
in the rotation. But if the polarization is partial, then there is a lot of
free (randomly orientated) vortex filaments that contribute to mutual
friction and do not contribute to the Feynman rule (\ref{rot via L}). In
addition, these chaotic lines interact with polarized lines, thereby
destroying the polarization and, correspondingly, the quasi-bundle structure.

There are other examples of observation of vortex bundles (in numerical
works), when they are artificially prepared structure, or are initiated by
eddies of normal component (see, e.g. \cite{Samuels1993},\cite{Samuels1993}%
). However, there are works in the literature, in which it is stated that
even if the vortex bundles are artificially created, they can be destroyed
rather soon. For instance, G. Volovik \cite{Volovik2004}) have shown that at
low temperatures, where the mutual friction is small, the existence of the
bundles is impossible. They should melt, changing into a highly irregular
structure. Other example is a series of numerical simulations by Kivotides
\cite{Kivotides2011}, \cite{Kivotides2012},\cite{Kivotides2014},\cite%
{Kivotides2018}), who studied the exact (not HVBK) dynamics of quantum
vortices in the turbulent flows (at finite temperature) and concluded "that
the results do not show that a turbulent normal-fluid with a Kolmogorov
energy spectrum induces superfluid vortex bundles in the superfluid".\textsl{%
\ }In paper \cite{Kivotides2014} Kivotides reported about observation of
clusters with weakly polarized vortex lines and associated Kolmogorov - type
spectra $E(k)\propto k^{5/3}$. To some measure this is expected result since
the appearance of the Kolmogorov spectrum requires formation of coherent
structures. However, the partial polarization also prevents the use of
closure $\nabla \times v_{s}=\kappa L$\ in the mutual friction, since ALL
the lines contribute into friction, and partial polarization (if any)
includes A SMALL FRACTION of the total vortex line density.

\ In this regard, it seems appropriate to discuss question of relation of
the vortex bundle arrangement and the Kolmogorov-type spectrum. For the
uniform array of vortex filaments the coarse grained velocity field is $%
\mathbf{v(r)=\Omega \times r}$. Accordingly the Fourier transform $\mathbf{%
v(k)}$\ scales as $1/k^{-3}$. This implies that two-dimensional spectrum $E(%
\mathbf{k})=$ $dE/d^{2}\mathbf{k}$\ should behave as $1/k^{-6}$\ \ and the
isotropic spectrum $E(k)$\ depends on absolute value of the wave vector $k$
as $E(k)\varpropto 1/k^{5}$. Thus, we state that the uniform vortex bundles
do not generate the Kolmogorov type spectra. It can be shown that the
nonuniform vortex array with the distribution of density of vortex filaments
$n(r)=\Delta N/\Delta r\varpropto 1/r^{-2/3}$\ does generate Kolmogorov type
spectra $E(k)\varpropto 1/k^{5/3}$\ (see \cite{Kozik2009}, \cite%
{Nemirovskii2015a}).

Of course, for very strong mutual friction vortex filaments can be
completely trapped by the eddies of normal component and follow the dynamics
of normal fluid. In fact, in this situation the coarse- grained
hydrodynamics become the one-fluid dynamics, when both components move
together.

There are many physical mechanisms that result in the destruction of the
regular vortex bundle structure. The apparent source of the destruction of
the bundle structures is the various reconnections. Thus, as demonstrated in
the paper by Kursa, Bajer \& Lipniacki, \cite{Kursa2011}, and in work by
Kerr \cite{Kerr2011} even a single reconnection results in a cascade of
vortex loops of various sizes being chaotically radiated from the
reconnection point. Clearly these propagating loops collide with the lines
composing bundles, triggering new reconnections, and developing an
avalanche-like randomization.

Some authors(see e.g. \cite{Alamri2008},\cite{Baggaley2012c}), based on
their numerical results, claim that the vortex bundles are the robust
structures with respect to reconnection of two adjacent bundles. This
conclusion, however, concerns the situation when the different bundles have
the same structure (N strands in each bundle)\textsl{. } In quantum fluids
full reconnection between bundles carrying different numbers of threads is
not possible for topological reasons, and the residual structure, analogous
to the "bridging" in classical hydrodynamics, should accompany the collision
of such bundles (see e.g. papers \cite{Melander1989}, \cite{Zabusky1989},
\cite{Kida1991}, \cite{Boratav1992}). The analog of classical "bridging"
leads to the randomization and violation of the structure of the bundles and
to the creation of vortex loops. Also, the long-range interaction between
vortex filaments in the bundles and the "external" vortices also destroy
regular array due to the action of tidal forces.

As for the filaments inside bundle, the direct reconnection event for them
is impossible, since for the reconnection the approaching vortices must be
antiparallel. Reorganization of lines destroys the parallel array of vortex
filaments.

Similarly, the processes of emission and re-absorption of the ring by the
vortices (the "anti-bottleneck", proposed by Svistunov \cite%
{Svistunov1995,Kozik2009}) should also lead to the fragmentation of the
regular arrays and the appearance of chaotic loops.

There are also experimental results\ which do questionable the idea of
bundles. Thus, in experiments by Roche et al. \cite{Roche2007}, \ and by
Bradley et al.\cite{Bradley2008},\ it was observed that the spectrum of the
fluctuation of the VLD $\mathcal{L}$ is compatible with a $-5/3$ power law.
This contradicts the idea of the bundles structure, since the spectrum of
the vorticity (and, correspondingly, of the VLD $\mathcal{L}$ (via Eq. (\ref%
{rot via L}))) should scale as $1/3$ power law, which,indeed was observed in
paper by Baggaley \cite{Baggaley2012a}.

One more objection to using HVBK approach is that the Feynman rule (\ref{rot
via L}) is applicable only for stationary situations. It is not clear its
validity for transient processes that take place in highly fluctuating
turbulent flows.

Resuming this subsection we would like to stress, that the bundle structure
of the quasi-classical quantum turbulence, although clear and transparent,
is not strongly confirmed . The regular vortex bundles, even if they
spontaneously appear, are extremely unstable structures,which can be easily
destroyed.

\subsection{\textsl{\ }Where is it from?}

This is a somewhat mysterious question: how did at all the HVBK method,
designed to a rotational or two-dimensional case, become used for
three-dimensional turbulent flows? Trying to find the origin I analyzed a
large mass of literary sources. The most frequent references are to the
papers by Sonin \cite{Sonin1987} and \cite{Hills1977a}. But these links are
absolutely irrelevant, since the authors definitely wrote that they work
with rotating helium. Probably, the one of the first papers in which the use
of this method for three-dimensional turbulent flows is discussed is the
work of Holm \cite{Holm2001}. It is interesting, however, that he started
paper with the text "Recent experiments establish the
Hall-Vinen-Bekarevich-Khalatnikov (HVBK) equations as a leading model for
describing superfluid Helium turbulence. See Nemirovskii and Fiszdon [1995]
and Donnelly [1999] for authoritative reviews."

But R. Donnelly \cite{Donnelly1999} discussed HVBK approach namely for
rotating helium. As for my (with W. Fiszdon) paper \cite{Nemirovskii1995} on
superfluid turbulence, then, firstly, there was no mention on HVBK theory at
all, and, secondly, I generally opposed to this method to treat
three-dimensional quantum turbulence.

Thus, the origin of the idea of the using a pure rotational HVBK
approximation for a three-dimensional turbulent flows is rather vague.

Resuming this Section, I can state the HVBK ansatz $\nabla \times \mathbf{v}%
_{s}=\kappa \mathcal{L}$ is in general unfounded in the three-dimensional
case, therefore, works using this approach are questionable, and the
corresponding results are not reliable.

\section{Other ways to treat the vortex line density}

The main attractive advantage in HVBK approach was to rid out of vortex line
density $\mathcal{L}$ in equations of motions (\ref{equa-Vn}) - (\ref%
{Fns-media}). It seems,however, that the question of "elimination" the
vortex line density $\mathcal{L}(\mathbf{r},t)$ should be solved in a
fundamentally different way. We have not to "eliminate" the quantity $%
\mathcal{L}(\mathbf{r},t)$, but on the contrary, to include it into
consideration as an independent and equipollent variable. Correspondingly we
have to consider the problem, in which there are three independent variables
- velocities $\mathbf{v}_{n}(\mathbf{r},t)$ and $\mathbf{v}_{s}(\mathbf{r}%
,t) $, and vortex line density $\mathcal{L}(\mathbf{r},t)$. We also can add
the density field $\rho (\mathbf{r},t),$~as\ well as the entropy field $S(%
\mathbf{r},t)$ if it is pertinent. In this way, however, we need an
additional independent equation for the temporal and spacial evolution of
quantity $\mathcal{L}(\mathbf{r},t)$.

This is a particular task that requires a lot of efforts. Derivation of such
an equation, certainly, depends on the type of flow, such as counterflow,
co-flow, flow past objects, unsteady rotation, etc. In fact, so far the
corresponding equation exists only for the case of counterflow,\ this is the
famous Vinen equation (or some modernized versions of this equation).
Although, there are some problems with this equation (see e.g. Sec. IV in
\cite{Nemirovskii2018e}), it works well for hydrodynamic e.g. for acoustic
or engineering problems. It is important to stress that the construction of
a theory of the evolution of three fields ($\mathbf{v}_{n}$, $\mathbf{v}_{s}$
and $\mathcal{L}(\mathbf{r},t)$) is not an automatic addition of Vinen
equation to the Landau two-fluid hydrodynamics. It is a more involved
procedure, since all variables (energy, entropy, etc.) change in the
presence of the vortex tangle. This self-consistent procedure was
implemented in \cite{Nemirovskii1983}, it is called Hydrodynamics of
Superfluid Turbulence (HST).

This theory was successful, it explained a lot of experimental results on
the nonlinear acoustics of the first and second sounds, the evolution of
strong heat pulses, the formation of shock waves etc. It was also
successfully applied to the problems of unsteady heat transfer and boiling
of He II (\cite{Nemirovskii1995},\cite{Jou2005},\cite{Kondaurova2017}).
These examples demonstrate that the way to treat the vortex line density $%
\mathcal{L}(\mathbf{r},t)$ \ as an additional and equipollent variable seems
to be productive and fruitful.

Unfortunately, for other types of flows there is no theory describing
temporal - spatial evolution of the vortex line density, and there is no
ideas (similar to the Feynman qualitative scenario) how to obtain the
according equation. Vinen's equation reflects the fact that the vortex line
density $\mathcal{L}(\mathbf{r},t)$ grows due to the relative velocity $%
\mathbf{v}_{n}-\mathbf{v}_{s}$ and attenuates, probably, due to the cascade
like breaking down of vortex loops, described by Feynman \cite{Feynman1955}.
That is good guideline how to develop an appropriate theory for any flow,
involving, of course, some auxiliary speculations.

Beside of introduction of the vortex line density into the coarse-grained
hydrodynamics of superfluid turbulence, there was one more, crude and
simplified way, which had been applied on the early stages of research on
the superfluid turbulence. It was the use the Gorter - Mellink formula for
mutual friction which immediately follows, from the Vinen theory

\begin{equation}
\mathbf{F}_{mf}\propto A(T)(v_{n}-v_{s})^{2}(\mathbf{v}_{n}-\mathbf{v}_{s}).
\label{GM}
\end{equation}%
Here $A(T)$ is the Gorter - Mellink. In fact, in this formula it was used
the ansatz\textsl{\ }$\mathcal{L}\propto (v_{n}-v_{s})^{2}$, well known in
the theory of quantum turbulence The use of equation (\ref{GM}) also
"resolves" the problem of elimination the vortex line density $\mathcal{L}%
(r,t)$.

\section{Conclusion}

The two approaches how to investigate turbulent flow in superfluids - HVBK
and \ HST have been described in the paper. In the first one, the vortex
line density $\mathcal{L}(\mathbf{r},t)$, crucial for the whole dynamics is
straightforwardly excluded from equations of motion (\ref{equa-Vn}) and (\ref%
{equa-Vs}) with the use of \ Feynman rule (\ref{rot via L}). In the second
approach, the variable $\mathcal{L}(r,t)$ is considered as an additional
independent variable obeying the according evolution equation.

In the present paper it had been argued that the HVBK ansatz $\nabla \times
\mathbf{v}_{s}=\kappa \mathcal{L}$ is suitable only for rotating cases and
fails in three-dimensional situations. The attempts to justify this
procedure give rise to a whole scientific direction (trend), which asserts
that the vortex tangle in quantum turbulence is composed of the so called
vortex bundles containing a set of near parallel lines. In the paper, I put
a number of arguments criticizing the conception of the vortex bundle
structure. References to the fact that in some numerical works a partial
polarization of vortex filaments had been observed cannot be considered as
justification for using the ansatz $\nabla \times \mathbf{v}_{s}=\kappa
\mathcal{L}$, since ALL the lines contribute into friction, and polarization
(if any) includes a small fraction of the total vortex line density.

Furthermore, it is of great concern that the concept of vortex bundles has
gone beyond the coarse-grained hydrodynamics of superfluid turbulence and
often serves as the basis for other (more subtle) aspects of the theory of
quantum turbulence. This seems counterproductive, since after the pioneering
works of Feynman, Vinen, Donnelly, Schwartz and others, it was customary to
present a vortex tangle as a set of stochastic loops with rich and diverse
dynamics. These loops are subject to large deformations (due to highly
nonlinear dynamics), they reconnect with each other and with the wall, split
and merge, creating a lot of daughter loops. This vision was observed in
numerous numerical simulations (see e.g. \cite{Schwarz1988},\cite{Aarts1994},%
\cite{Tsubota2000},\cite{Berloff2002}, \cite{Kondaurova2008},\cite%
{Kivotides2014}). This, let's say, Feynman-Vinen model is very different.
from the vortex bundle model, where almost the only possible dynamics of the
vortex filaments is the Kelvin waves evolution along the lines composing the
bundles.

In summary, the use of ansatz $\nabla \times \mathbf{v}_{s}=\kappa \mathcal{L%
}$, for the closure procedure for coupled Navier-Stokes equations (\ref%
{equa-Vn}) - (\ref{Fns-media}) in the 3D turbulent flow is not motivated and
would lead to unreliable results. And the commonly used the vortex bundle
model, which justifies the use of this method, is questionable and
unfounded. Moreover, the vortex bundle concept disavows the real structure
of the vortex tangle as a set of vortex loops, and prevents developing of an
adequate theory. We assert that the introduction of an additional
independent field $\mathcal{L}(r,t)$ to the classical two-fluid
hydrodynamics of supefluids is the only correct way to construct the
coarse-garined hydrodynamics of turbulent flows.

\section{Acknowledgements.}

The study on the Hall-Vinen-Bekarevich-Khalatnikov (HVBK) approach was
carried out under state contract with IT SB RAS (No.
17-117022850027-5), the study on the Hydrodynamics of
Superfluid Turbulence (HST) method  was financially supported by RFBR 
Russian Science Foundation (Project No. 18-08-00576).


\begin{thebibliography}{40}
\expandafter\ifx\csname natexlab\endcsname\relax\def\natexlab#1{#1}\fi
\expandafter\ifx\csname bibnamefont\endcsname\relax
  \def\bibnamefont#1{#1}\fi
\expandafter\ifx\csname bibfnamefont\endcsname\relax
  \def\bibfnamefont#1{#1}\fi
\expandafter\ifx\csname citenamefont\endcsname\relax
  \def\citenamefont#1{#1}\fi
\expandafter\ifx\csname url\endcsname\relax
  \def\url#1{\texttt{#1}}\fi
\expandafter\ifx\csname urlprefix\endcsname\relax\def\urlprefix{URL }\fi
\providecommand{\bibinfo}[2]{#2}
\providecommand{\eprint}[2][]{\url{#2}}

\bibitem[{\citenamefont{Gorter and Mellink}(1949)}]{Gorter1949}
\bibinfo{author}{\bibfnamefont{C.~J.} \bibnamefont{Gorter}} \bibnamefont{and}
  \bibinfo{author}{\bibfnamefont{J.~H.} \bibnamefont{Mellink}},
  \bibinfo{journal}{Physica} \textbf{\bibinfo{volume}{15}}, \bibinfo{pages}{285
  } (\bibinfo{year}{1949}).

\bibitem[{\citenamefont{Feynman}(1955)}]{Feynman1955}
\bibinfo{author}{\bibfnamefont{R.~P.} \bibnamefont{Feynman}},
  \emph{\bibinfo{title}{Progress in Low Temperature Physics, Vol. 1}}
  (\bibinfo{publisher}{North-Holland, Amsterdam}, \bibinfo{year}{1955}),
  p.~\bibinfo{pages}{17}.

\bibitem[{\citenamefont{Schwarz}(1988)}]{Schwarz1988}
\bibinfo{author}{\bibfnamefont{K.~W.} \bibnamefont{Schwarz}},
  \bibinfo{journal}{Phys. Rev. B} \textbf{\bibinfo{volume}{38}},
  \bibinfo{pages}{2398} (\bibinfo{year}{1988}).

\bibitem[{\citenamefont{Donnelly}(1991)}]{Donnelly1991}
\bibinfo{author}{\bibfnamefont{R.}~\bibnamefont{Donnelly}},
  \emph{\bibinfo{title}{Quantized Vortices in Helium II}}
  (\bibinfo{publisher}{Cambridge University Press, Cambridge, UK},
  \bibinfo{year}{1991}).

\bibitem[{\citenamefont{Khalatnikov}(1965)}]{Khalatnikov1965}
\bibinfo{author}{\bibfnamefont{I.~M.} \bibnamefont{Khalatnikov}},
  \emph{\bibinfo{title}{An Introduction to the Theory of Superfluidity}}
  (\bibinfo{publisher}{Benjamin, New York/Amsterdam}, \bibinfo{year}{1965}).

\bibitem[{\citenamefont{Nemirovskii and Lebedev}(1983)}]{Nemirovskii1983}
\bibinfo{author}{\bibfnamefont{S.}~\bibnamefont{Nemirovskii}} \bibnamefont{and}
  \bibinfo{author}{\bibfnamefont{V.}~\bibnamefont{Lebedev}},
  \bibinfo{journal}{Sov. Phys. JETP} \textbf{\bibinfo{volume}{57}},
  \bibinfo{pages}{1009} (\bibinfo{year}{1983}).

\bibitem[{\citenamefont{Nemirovskii and Fiszdon}(1995)}]{Nemirovskii1995}
\bibinfo{author}{\bibfnamefont{S.~K.} \bibnamefont{Nemirovskii}}
  \bibnamefont{and} \bibinfo{author}{\bibfnamefont{W.}~\bibnamefont{Fiszdon}},
  \bibinfo{journal}{Rev. Mod. Phys.} \textbf{\bibinfo{volume}{67}},
  \bibinfo{pages}{37} (\bibinfo{year}{1995}).

\bibitem[{\citenamefont{Sonin}(2016)}]{Sonin2016}
\bibinfo{author}{\bibfnamefont{E.~B.} \bibnamefont{Sonin}},
  \emph{\bibinfo{title}{Dynamics of Quantised Vortices in Superfluids}}
  (\bibinfo{publisher}{Cambridge University Press}, \bibinfo{year}{2016}).

\bibitem[{\citenamefont{Baggaley et~al.}(2012)\citenamefont{Baggaley, Barenghi,
  Shukurov, and Sergeev}}]{Baggaley2012a}
\bibinfo{author}{\bibfnamefont{A.~W.} \bibnamefont{Baggaley}},
  \bibinfo{author}{\bibfnamefont{C.~F.} \bibnamefont{Barenghi}},
  \bibinfo{author}{\bibfnamefont{A.}~\bibnamefont{Shukurov}}, \bibnamefont{and}
  \bibinfo{author}{\bibfnamefont{Y.~A.} \bibnamefont{Sergeev}},
  \bibinfo{journal}{EPL (Europhysics Letters)} \textbf{\bibinfo{volume}{98}},
  \bibinfo{pages}{26002} (\bibinfo{year}{2012}).

\bibitem[{\citenamefont{Sasa et~al.}(2011)\citenamefont{Sasa, Kano, Machida,
  L'vov, Rudenko, and Tsubota}}]{Sasa2011}
\bibinfo{author}{\bibfnamefont{N.}~\bibnamefont{Sasa}},
  \bibinfo{author}{\bibfnamefont{T.}~\bibnamefont{Kano}},
  \bibinfo{author}{\bibfnamefont{M.}~\bibnamefont{Machida}},
  \bibinfo{author}{\bibfnamefont{V.~S.} \bibnamefont{L'vov}},
  \bibinfo{author}{\bibfnamefont{O.}~\bibnamefont{Rudenko}}, \bibnamefont{and}
  \bibinfo{author}{\bibfnamefont{M.}~\bibnamefont{Tsubota}},
  \bibinfo{journal}{Phys. Rev. B} \textbf{\bibinfo{volume}{84}},
  \bibinfo{pages}{054525} (\bibinfo{year}{2011}).

\bibitem[{\citenamefont{Samuels}(1993)}]{Samuels1993}
\bibinfo{author}{\bibfnamefont{D.~C.} \bibnamefont{Samuels}},
  \bibinfo{journal}{Phys. Rev. B} \textbf{\bibinfo{volume}{47}},
  \bibinfo{pages}{1107} (\bibinfo{year}{1993}).

\bibitem[{\citenamefont{Volovik}(2004)}]{Volovik2004}
\bibinfo{author}{\bibfnamefont{G.}~\bibnamefont{Volovik}},
  \bibinfo{journal}{Journal of Low Temperature Physics}
  \textbf{\bibinfo{volume}{136}}, \bibinfo{pages}{309} (\bibinfo{year}{2004}),
  ISSN \bibinfo{issn}{1573-7357}.

\bibitem[{\citenamefont{Kivotides}(2011)}]{Kivotides2011}
\bibinfo{author}{\bibfnamefont{D.}~\bibnamefont{Kivotides}},
  \bibinfo{journal}{Journal of Fluid Mechanics} \textbf{\bibinfo{volume}{668}},
  \bibinfo{pages}{58} (\bibinfo{year}{2011}).

\bibitem[{\citenamefont{Kivotides}(2012)}]{Kivotides2012}
\bibinfo{author}{\bibfnamefont{D.}~\bibnamefont{Kivotides}}, in
  \emph{\bibinfo{booktitle}{New Approaches in Modeling Multiphase Flows and
  Dispersion in Turbulence, Fractal Methods and Synthetic Turbulence}}, edited
  by \bibinfo{editor}{\bibfnamefont{F.}~\bibnamefont{Nicolleau}},
  \bibinfo{editor}{\bibfnamefont{C.}~\bibnamefont{Cambon}},
  \bibinfo{editor}{\bibfnamefont{J.-M.} \bibnamefont{Redondo}},
  \bibinfo{editor}{\bibfnamefont{J.}~\bibnamefont{Vassilicos}},
  \bibinfo{editor}{\bibfnamefont{M.}~\bibnamefont{Reeks}}, \bibnamefont{and}
  \bibinfo{editor}{\bibfnamefont{A.}~\bibnamefont{Nowakowski}}
  (\bibinfo{publisher}{Springer Netherlands}, \bibinfo{year}{2012}),
  vol.~\bibinfo{volume}{18} of \emph{\bibinfo{series}{ERCOFTAC Series}}, pp.
  \bibinfo{pages}{1--8}, ISBN \bibinfo{isbn}{978-94-007-2505-8}.

\bibitem[{\citenamefont{Kivotides}(2014)}]{Kivotides2014}
\bibinfo{author}{\bibfnamefont{D.}~\bibnamefont{Kivotides}},
  \bibinfo{journal}{Physics of Fluids (1994-present)}
  \textbf{\bibinfo{volume}{26}}, \bibinfo{pages}{105105}
  (\bibinfo{year}{2014}).

\bibitem[{\citenamefont{Kivotides}(2018)}]{Kivotides2018}
\bibinfo{author}{\bibfnamefont{D.}~\bibnamefont{Kivotides}},
  \bibinfo{journal}{Physics Letters A} \textbf{\bibinfo{volume}{382}},
  \bibinfo{pages}{1481 } (\bibinfo{year}{2018}), ISSN
  \bibinfo{issn}{0375-9601}.

\bibitem[{\citenamefont{Kozik and Svistunov}(2009)}]{Kozik2009}
\bibinfo{author}{\bibfnamefont{E.}~\bibnamefont{Kozik}} \bibnamefont{and}
  \bibinfo{author}{\bibfnamefont{B.}~\bibnamefont{Svistunov}},
  \bibinfo{journal}{Journal of Low Temperature Physics}
  \textbf{\bibinfo{volume}{156}}, \bibinfo{pages}{215} (\bibinfo{year}{2009}).

\bibitem[{\citenamefont{Nemirovskii}(2015)}]{Nemirovskii2015a}
\bibinfo{author}{\bibfnamefont{S.~K.} \bibnamefont{Nemirovskii}},
  \bibinfo{journal}{Low Temperature Physics} \textbf{\bibinfo{volume}{41}},
  \bibinfo{pages}{608} (\bibinfo{year}{2015}).

\bibitem[{\citenamefont{Kursa et~al.}(2011)\citenamefont{Kursa, Bajer, and
  Lipniacki}}]{Kursa2011}
\bibinfo{author}{\bibfnamefont{M.}~\bibnamefont{Kursa}},
  \bibinfo{author}{\bibfnamefont{K.}~\bibnamefont{Bajer}}, \bibnamefont{and}
  \bibinfo{author}{\bibfnamefont{T.}~\bibnamefont{Lipniacki}},
  \bibinfo{journal}{Phys. Rev. B} \textbf{\bibinfo{volume}{83}},
  \bibinfo{pages}{014515} (\bibinfo{year}{2011}).

\bibitem[{\citenamefont{Kerr}(2011)}]{Kerr2011}
\bibinfo{author}{\bibfnamefont{R.~M.} \bibnamefont{Kerr}},
  \bibinfo{journal}{Phys. Rev. Lett.} \textbf{\bibinfo{volume}{106}},
  \bibinfo{pages}{224501} (\bibinfo{year}{2011}).

\bibitem[{\citenamefont{Alamri et~al.}(2008)\citenamefont{Alamri, Youd, and
  Barenghi}}]{Alamri2008}
\bibinfo{author}{\bibfnamefont{S.~Z.} \bibnamefont{Alamri}},
  \bibinfo{author}{\bibfnamefont{A.~J.} \bibnamefont{Youd}}, \bibnamefont{and}
  \bibinfo{author}{\bibfnamefont{C.~F.} \bibnamefont{Barenghi}},
  \bibinfo{journal}{Phys. Rev. Lett.} \textbf{\bibinfo{volume}{101}},
  \bibinfo{pages}{215302} (\bibinfo{year}{2008}).

\bibitem[{\citenamefont{Baggaley}(2012)}]{Baggaley2012c}
\bibinfo{author}{\bibfnamefont{A.~W.} \bibnamefont{Baggaley}},
  \bibinfo{journal}{Physics of Fluids} \textbf{\bibinfo{volume}{24}},
  \bibinfo{eid}{055109} (pages~\bibinfo{numpages}{9}) (\bibinfo{year}{2012}).

\bibitem[{\citenamefont{Melander and Hussain}(1989)}]{Melander1989}
\bibinfo{author}{\bibfnamefont{M.~V.} \bibnamefont{Melander}} \bibnamefont{and}
  \bibinfo{author}{\bibfnamefont{F.}~\bibnamefont{Hussain}},
  \bibinfo{journal}{Phys. Fluids} \textbf{\bibinfo{volume}{A 1}},
  \bibinfo{pages}{633} (\bibinfo{year}{1989}).

\bibitem[{\citenamefont{Zabusky and Melander}(1989)}]{Zabusky1989}
\bibinfo{author}{\bibfnamefont{N.}~\bibnamefont{Zabusky}} \bibnamefont{and}
  \bibinfo{author}{\bibfnamefont{M.}~\bibnamefont{Melander}},
  \bibinfo{journal}{Physica D: Nonlinear Phenomena}
  \textbf{\bibinfo{volume}{37}}, \bibinfo{pages}{555 } (\bibinfo{year}{1989}),
  ISSN \bibinfo{issn}{0167-2789}.

\bibitem[{\citenamefont{Kida et~al.}(1991)\citenamefont{Kida, Takaoka, and
  Hussain}}]{Kida1991}
\bibinfo{author}{\bibfnamefont{S.}~\bibnamefont{Kida}},
  \bibinfo{author}{\bibfnamefont{M.}~\bibnamefont{Takaoka}}, \bibnamefont{and}
  \bibinfo{author}{\bibfnamefont{F.}~\bibnamefont{Hussain}},
  \bibinfo{journal}{Journal of Fluid Mechanics} \textbf{\bibinfo{volume}{230}},
  \bibinfo{pages}{583} (\bibinfo{year}{1991}).

\bibitem[{\citenamefont{Boratav et~al.}(1992)\citenamefont{Boratav, Pelz, and
  Zabusky}}]{Boratav1992}
\bibinfo{author}{\bibfnamefont{O.}~\bibnamefont{Boratav}},
  \bibinfo{author}{\bibfnamefont{R.}~\bibnamefont{Pelz}}, \bibnamefont{and}
  \bibinfo{author}{\bibfnamefont{N.}~\bibnamefont{Zabusky}},
  \bibinfo{journal}{Phys. Fluids A} \textbf{\bibinfo{volume}{4}},
  \bibinfo{pages}{581} (\bibinfo{year}{1992}).

\bibitem[{\citenamefont{Svistunov}(1995)}]{Svistunov1995}
\bibinfo{author}{\bibfnamefont{B.~V.} \bibnamefont{Svistunov}},
  \bibinfo{journal}{Phys. Rev. B} \textbf{\bibinfo{volume}{52}},
  \bibinfo{pages}{3647} (\bibinfo{year}{1995}).

\bibitem[{\citenamefont{Roche et~al.}(2007)\citenamefont{Roche, Diribarne,
  Didelot, Francais, Rousseau, and Willaime}}]{Roche2007}
\bibinfo{author}{\bibfnamefont{P.-E.} \bibnamefont{Roche}},
  \bibinfo{author}{\bibfnamefont{P.}~\bibnamefont{Diribarne}},
  \bibinfo{author}{\bibfnamefont{T.}~\bibnamefont{Didelot}},
  \bibinfo{author}{\bibfnamefont{O.}~\bibnamefont{Francais}},
  \bibinfo{author}{\bibfnamefont{L.}~\bibnamefont{Rousseau}}, \bibnamefont{and}
  \bibinfo{author}{\bibfnamefont{H.}~\bibnamefont{Willaime}},
  \bibinfo{journal}{EPL (Europhysics Letters)} \textbf{\bibinfo{volume}{77}},
  \bibinfo{pages}{66002} (\bibinfo{year}{2007}).

\bibitem[{\citenamefont{Bradley et~al.}(2008)\citenamefont{Bradley, Fisher,
  Gu\'enault, Haley, O'Sullivan, Pickett, and Tsepelin}}]{Bradley2008}
\bibinfo{author}{\bibfnamefont{D.~I.} \bibnamefont{Bradley}},
  \bibinfo{author}{\bibfnamefont{S.~N.} \bibnamefont{Fisher}},
  \bibinfo{author}{\bibfnamefont{A.~M.} \bibnamefont{Gu\'enault}},
  \bibinfo{author}{\bibfnamefont{R.~P.} \bibnamefont{Haley}},
  \bibinfo{author}{\bibfnamefont{S.}~\bibnamefont{O'Sullivan}},
  \bibinfo{author}{\bibfnamefont{G.~R.} \bibnamefont{Pickett}},
  \bibnamefont{and} \bibinfo{author}{\bibfnamefont{V.}~\bibnamefont{Tsepelin}},
  \bibinfo{journal}{Phys. Rev. Lett.} \textbf{\bibinfo{volume}{101}},
  \bibinfo{pages}{065302} (\bibinfo{year}{2008}).

\bibitem[{\citenamefont{Sonin}(1987)}]{Sonin1987}
\bibinfo{author}{\bibfnamefont{E.~B.} \bibnamefont{Sonin}},
  \bibinfo{journal}{Rev. Mod. Phys.} \textbf{\bibinfo{volume}{59}},
  \bibinfo{pages}{87} (\bibinfo{year}{1987}).

\bibitem[{\citenamefont{Hills and Roberts}(1977)}]{Hills1977a}
\bibinfo{author}{\bibfnamefont{R.~N.} \bibnamefont{Hills}} \bibnamefont{and}
  \bibinfo{author}{\bibfnamefont{P.~H.} \bibnamefont{Roberts}},
  \bibinfo{journal}{Archive for Rational Mechanics and Analysis}
  \textbf{\bibinfo{volume}{66}}, \bibinfo{pages}{43} (\bibinfo{year}{1977}),
  ISSN \bibinfo{issn}{1432-0673}.

\bibitem[{\citenamefont{Holm}(2001)}]{Holm2001}
\bibinfo{author}{\bibfnamefont{D.~D.} \bibnamefont{Holm}}, in
  \emph{\bibinfo{booktitle}{Quantized Vortex Dynamics}}
  (\bibinfo{publisher}{Springer}, \bibinfo{year}{2001}).

\bibitem[{\citenamefont{Donnelly}(1999)}]{Donnelly1999}
\bibinfo{author}{\bibfnamefont{R.~J.} \bibnamefont{Donnelly}},
  \bibinfo{journal}{Journal of Physics: Condensed Matter}
  \textbf{\bibinfo{volume}{11}}, \bibinfo{pages}{7783} (\bibinfo{year}{1999}).

\bibitem[{\citenamefont{Nemirovskii}(2018)}]{Nemirovskii2018e}
\bibinfo{author}{\bibfnamefont{S.~K.} \bibnamefont{Nemirovskii}},
  \bibinfo{journal}{Phys. Rev. B} \textbf{\bibinfo{volume}{97}},
  \bibinfo{pages}{134511} (\bibinfo{year}{2018}),
  \urlprefix\url{https://link.aps.org/doi/10.1103/PhysRevB.97.134511}.

\bibitem[{\citenamefont{Jou and Mongiov\`\i}(2005)}]{Jou2005}
\bibinfo{author}{\bibfnamefont{D.}~\bibnamefont{Jou}} \bibnamefont{and}
  \bibinfo{author}{\bibfnamefont{M.~S.} \bibnamefont{Mongiov\`\i}},
  \bibinfo{journal}{Phys. Rev. B} \textbf{\bibinfo{volume}{72}},
  \bibinfo{pages}{144517} (\bibinfo{year}{2005}).

\bibitem[{\citenamefont{Kondaurova et~al.}(2017)\citenamefont{Kondaurova,
  Efimov, and Tsoi}}]{Kondaurova2017}
\bibinfo{author}{\bibfnamefont{L.}~\bibnamefont{Kondaurova}},
  \bibinfo{author}{\bibfnamefont{V.}~\bibnamefont{Efimov}}, \bibnamefont{and}
  \bibinfo{author}{\bibfnamefont{A.}~\bibnamefont{Tsoi}},
  \bibinfo{journal}{Journal of Low Temperature Physics}
  \textbf{\bibinfo{volume}{187}}, \bibinfo{pages}{80} (\bibinfo{year}{2017}),
  ISSN \bibinfo{issn}{1573-7357}.

\bibitem[{\citenamefont{Aarts and de~Waele}(1994)}]{Aarts1994}
\bibinfo{author}{\bibfnamefont{R.~G. K.~M.} \bibnamefont{Aarts}}
  \bibnamefont{and} \bibinfo{author}{\bibfnamefont{A.~T. A.~M.}
  \bibnamefont{de~Waele}}, \bibinfo{journal}{Phys. Rev. B}
  \textbf{\bibinfo{volume}{50}}, \bibinfo{pages}{10069} (\bibinfo{year}{1994}).

\bibitem[{\citenamefont{Tsubota et~al.}(2000)\citenamefont{Tsubota, Araki, and
  Nemirovskii}}]{Tsubota2000}
\bibinfo{author}{\bibfnamefont{M.}~\bibnamefont{Tsubota}},
  \bibinfo{author}{\bibfnamefont{T.}~\bibnamefont{Araki}}, \bibnamefont{and}
  \bibinfo{author}{\bibfnamefont{S.~K.} \bibnamefont{Nemirovskii}},
  \bibinfo{journal}{Phys. Rev. B} \textbf{\bibinfo{volume}{62}},
  \bibinfo{pages}{11751} (\bibinfo{year}{2000}).

\bibitem[{\citenamefont{Berloff and Svistunov}(2002)}]{Berloff2002}
\bibinfo{author}{\bibfnamefont{N.~G.} \bibnamefont{Berloff}} \bibnamefont{and}
  \bibinfo{author}{\bibfnamefont{B.~V.} \bibnamefont{Svistunov}},
  \bibinfo{journal}{Phys. Rev. A} \textbf{\bibinfo{volume}{66}},
  \bibinfo{pages}{013603} (\bibinfo{year}{2002}).

\bibitem[{\citenamefont{Kondaurova et~al.}(2008)\citenamefont{Kondaurova,
  Andryuschenko, and Nemirovskii}}]{Kondaurova2008}
\bibinfo{author}{\bibfnamefont{L.}~\bibnamefont{Kondaurova}},
  \bibinfo{author}{\bibfnamefont{V.}~\bibnamefont{Andryuschenko}},
  \bibnamefont{and}
  \bibinfo{author}{\bibfnamefont{S.}~\bibnamefont{Nemirovskii}},
  \bibinfo{journal}{Journal of Low Temperature Physics}
  \textbf{\bibinfo{volume}{150}}, \bibinfo{pages}{415} (\bibinfo{year}{2008}).

\end{thebibliography}

\end{document}